\documentclass[11pt]{article}

\usepackage{amsmath,amsfonts,amssymb,bbm,epsfig}
\usepackage[textwidth=14.7cm,top=30mm,bottom=35mm]{geometry}
\usepackage{tabularx}
\usepackage{braket}

\numberwithin{equation}{section} 

\renewcommand\slash[1]{\not \! #1}

\begin{document}

 \newcolumntype{L}[1]{>{\raggedright\arraybackslash}p{#1}}
 \newcolumntype{C}[1]{>{\centering\arraybackslash}p{#1}}
 \newcolumntype{R}[1]{>{\raggedleft\arraybackslash}p{#1}}

\def\be{\begin{equation}}
\def\ee{\end{equation}}
 \newcommand{\ba}{\begin{eqnarray}}
 \newcommand{\ea}{\end{eqnarray}}
\def\rarr{\rightarrow}
\def\nn{\nonumber}
\def\fr{\frac}
\renewcommand\slash[1]{\not \! #1}
\newcommand\qs{\!\not \! q}
\def\del{\partial}
\def\gam{\gamma}
\newcommand\vphi{\varphi}
\def\tr{\mbox{tr}\,}
\newcommand\nin{\noindent}

\def\del{\partial}
\def\pbar{\bar{p}}

\newcommand{\Pom}{\mathbb{P}}
\newcommand{\Ode}{\mathbb{O}}
\newcommand{\Reg}{\mathbb{R}}

\renewcommand\slash[1]{\not \! #1}

\newcommand{\bk}{\mbox{\boldmath $k$}}
\newcommand{\bea}{\mbox{\boldmath $e_{1}$}}
\newcommand{\beb}{\mbox{\boldmath $e_{2}$}}
\newcommand{\bec}{\mbox{\boldmath $e_{3}$}}

\begin{titlepage}
\begin{flushright}
\end{flushright}
\vspace{0.6cm}
\begin{center}
\boldmath
{\LARGE{\bf Helicity in Proton-Proton Elastic Scattering}}\\[.2cm]
{\LARGE{\bf and the Spin Structure of the Pomeron}}\\
\unboldmath
\end{center}
\vspace{0.6cm}
\begin{center}
{\bf \Large
Carlo Ewerz\,$^{a,b,c,1}$, Piotr Lebiedowicz\,$^{d,2}$,\\
Otto Nachtmann\,$^{a,3}$, Antoni Szczurek\,$^{d,4,\#}$}
\end{center}
\vspace{.2cm}
\begin{center}
$^a$
{\sl
Institut f\"ur Theoretische Physik, Universit\"at Heidelberg\\
Philosophenweg 16, D-69120 Heidelberg, Germany}
\\[.5cm]
$^b$
{\sl
ExtreMe Matter Institute EMMI, GSI Helmholtzzentrum f\"ur Schwerionenforschung\\
Planckstra{\ss}e 1, D-64291 Darmstadt, Germany}
\\[.5cm]
$^c$
{\sl
Frankfurt Institute for Advanced Studies\\
Ruth-Moufang-Stra{\ss}e 1, D-60438 Frankfurt, Germany
}
\\[.5cm]
$^d$
{\sl
Institute of Nuclear Physics, Polish Academy of Sciences\\ 
Radzikowskiego 152, PL-31-342 Krak{\'o}w, Poland
}
\end{center}                                                                
\vfill
\begin{abstract}
\noindent
We discuss different models for the spin structure of the nonperturbative 
pomeron: scalar, vector, and rank-2 symmetric tensor. 
The ratio of single-helicity-flip to helicity-conserving amplitudes 
in polarised high-energy proton-proton elastic scattering, known as the complex 
$r_5$ parameter, is calculated for these models. We compare our results to 
experimental data from the STAR experiment. 
We show that the spin-0 (scalar) pomeron model is clearly excluded by the 
data, while the vector pomeron is inconsistent with the rules of quantum field 
theory. The tensor pomeron is found to be perfectly consistent with the STAR data. 
\vfill
\end{abstract}
\vspace{5em}
\hrule width 5.cm
\vspace*{.5em}
{\small \noindent
$^1$ email: C.Ewerz@thphys.uni-heidelberg.de\\
$^2$ email: Piotr.Lebiedowicz@ifj.edu.pl\\
$^3$ email: O.Nachtmann@thphys.uni-heidelberg.de\\
$^4$ email: Antoni.Szczurek@ifj.edu.pl\\
$^\#$ Also at University of Rzesz\'ow, PL-35-959 Rzesz{\'o}w, Poland.
}
\end{titlepage}

\section{Introduction}
\label{sec:Introduction}
High-energy small-angle hadron-hadron scattering
is dominated by the exchange of the soft pomeron.
The nature of this pomeron has been discussed in a great number of articles; 
for reviews see, for instance, 
\cite{Donnachie:2002en,Caneschi:1989,Blois2005,Barone:2002cv}. 
It is clear that it has vacuum internal quantum numbers.
What is much less clear is the spin structure of the soft pomeron.
Indeed, the present authors have frequently been asked the following question: 
The pomeron has vacuum quantum numbers, 
should it then not also have spin zero?
However, a vector pomeron is widely used in the literature following 
\cite{Donnachie:1983hf,Donnachie:1985iz,Donnachie:1987gu}.
In \cite{Ewerz:2013kda} it was proposed to describe the soft pomeron
as an effective rank-2 symmetric tensor exchange. 
There, all reggeon exchanges with charge conjugation
$C = +1$ ($C = -1$) were described as effective
tensor (vector) exchanges and a large number of the couplings
of these objects to hadrons were determined from experimental data.
This tensor-pomeron model was then applied to various reactions in 
\cite{Lebiedowicz:2013ika,Bolz:2014mya,Lebiedowicz:2014bea,Lebiedowicz:2016ioh,Lebiedowicz:2016zka}.

The authoritative answer to the question for the spin structure of 
the soft pomeron should be given by experiment. 
In this article we want to show that experimental data 
on the helicity structure of small-$t$ proton-proton high-energy elastic scattering 
from the STAR experiment \cite{Adamczyk:2012kn} 
indeed give decisive information on the spin structure of the soft pomeron.

We emphasise that in the following we shall be interested
only in soft hadronic scattering where, according to standard wisdom,
perturbative QCD methods cannot be applied.

\section{Theoretical framework}
\label{sec:Theoretical_framework}

\begin{figure}
\begin{center}
\includegraphics[width=6.cm]{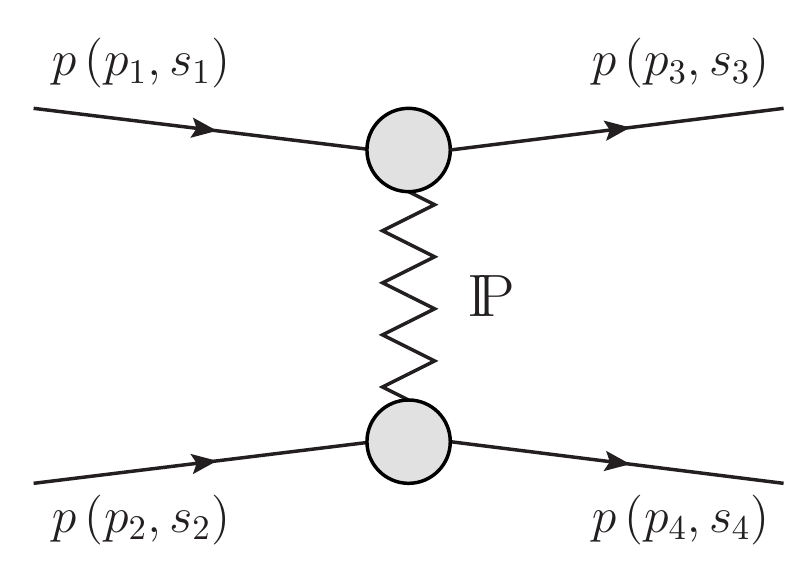}       
\caption{
Proton-proton elastic scattering via pomeron exchange.
\label{fig:pp_scattering}
}
\end{center}
\end{figure}

We will consider $pp$ elastic scattering
\begin{eqnarray}\label{pp_reaction}
p(p_{1},s_{1}) + p(p_{2},s_{2}) \longrightarrow
p(p_{3},s_{3}) + p(p_{4},s_{4}) \,,
\end{eqnarray}
where $p_{j}$ are the four-momenta and 
$s_{j} \in \{1/2,-1/2 \}$
the helicity indices, respectively.
The standard kinematic variables are 
\begin{equation}\label{kinematic_variables}
\begin{split}
&s = (p_{1} + p_{2})^{2} = (p_{3} + p_{4})^{2}\,,\\
&t = (p_{1} - p_{3})^{2} = (p_{2} - p_{4})^{2}\,,\\
&u = (p_{1} - p_{4})^{2} = (p_{2} - p_{3})^{2}\,.
\end{split}
\end{equation}
At high energies, $s \gg m_{p}^{2}$, $|t|$, the reaction (\ref{pp_reaction})
is dominated by pomeron exchange; see Fig.~\ref{fig:pp_scattering}. 
In the following all non-leading reggeon exchanges will be neglected.
For $\sqrt{s} \geqslant 200 \,\mbox{GeV}$ their contribution to the total cross section 
is only $\lesssim 1 \%$ \cite{Donnachie:2002en}. 

We shall test three hypotheses for the pomeron and the effective
pomeron-proton-proton ($\Pom p p$) vertex. We shall treat the pomeron 
either as a scalar, a vector, or a rank-2 symmetric tensor exchange.
For all three cases it turns out to be possible to adjust the effective pomeron propagators 
and the couplings in such a way that at high energies the helicity-conserving
$pp$ amplitudes have the standard form as given in the vector-pomeron 
model due to Donnachie and Landshoff; 
see \cite{Donnachie:1983hf,Donnachie:1985iz,Donnachie:1987gu} 
and \cite{Donnachie:2002en}. We will choose the parameters for the scalar 
and the tensor pomeron accordingly. 

\subsection{Tensor pomeron $\Pom_{T}$}
\label{sec:Tensor_pomeron}
Here we describe the pomeron, as discussed in \cite{Ewerz:2013kda},
as a symmetric, rank-two, tensor $\Pom_{T \mu \nu}(x)$
and its interaction with protons by coupling it to a tensor current $J_{T \mu \nu}(x)$,
\begin{equation}\label{tensor_pomeron_current}
\begin{split}
\mathcal{L}'_{T}(x) &= J_{T \mu \nu}(x) \,\Pom_{T}^{\mu \nu}(x)\,,\\
J_{T \mu \nu}(x) &=
- 3 \beta_{\Pom NN}
\frac{i}{2} \bar{\psi}_{p}(x)
\left[ 
\gamma_\mu \stackrel{\leftrightarrow}{\partial_{\nu} }
+ \gamma_\nu \stackrel{\leftrightarrow}{\partial_{\mu}}
- \frac{1}{2} g_{\mu\nu} \gamma^{\lambda}
\stackrel{\leftrightarrow}{\partial_{\lambda}} 
\right] \psi_{p}(x)\,;
\end{split}
\end{equation}
see (6.27) of \cite{Ewerz:2013kda}.
Here $\psi_{p}(x)$ is the proton field operator and
\begin{equation}\label{tensor_coupling}
\begin{split}
3 \beta_{\Pom NN} = 3 \times 1.87 \,{\rm GeV}^{-1}
\end{split}
\end{equation}
is the standard coupling constant describing the pomeron-nucleon interaction; 
see \cite{Donnachie:2002en,Ewerz:2013kda}.
From (\ref{tensor_pomeron_current}) we get the $\Pom_{T}pp$ vertex (see (3.43) 
of \cite{Ewerz:2013kda}) as\footnote{Note that from now on we will choose the 
orientation of the diagrams such that the $t$-channel is in horizontal and the 
$s$-channel in vertical direction.} 
\newline
\hspace*{1.8cm}\includegraphics[width=145pt]{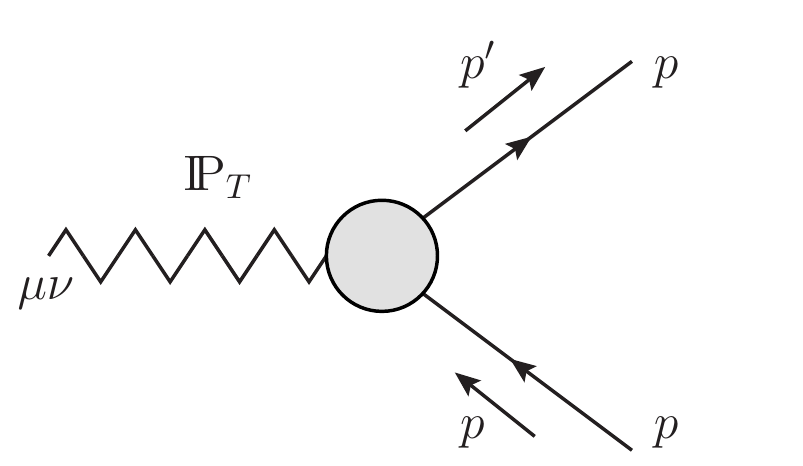} 
\begin{equation}\label{tensor_pomeron_vertex}
\begin{split}
i \Gamma_{\mu\nu}^{(\Pom_{T} pp)}(p',p) =
-i \,3 \beta_{\Pom NN} F_{1}[(p'-p)^2]
\left\{ \frac{1}{2} \left[ \gamma_\mu (p'+p)_\nu + \gamma_\nu (p'+p)_\mu \right] 
-\frac{1}{4} \, g_{\mu\nu} (\slash{p}' + \slash{p}) \right\}.
\end{split}
\end{equation}
A form factor, $F_{1}(t)$, has been introduced in the vertex (\ref{tensor_pomeron_vertex}).
Conventionally this is taken as the Dirac electromagnetic form factor of the proton,
following \cite{Donnachie:1983hf,Donnachie:1985iz,Donnachie:1987gu}. 
The normalisation is
\begin{equation}\label{F1_normalisation}
\begin{split}
F_{1}(0) = 1 \,.
\end{split}
\end{equation}
In the sequel the precise form of $F_{1}(t)$ will not be relevant. 

The ansatz for the effective $\Pom_{T}$ propagator is given in
(3.10) and (3.11) of \cite{Ewerz:2013kda} and reads
\newline
\hspace*{1.8cm}\includegraphics[width=125pt]{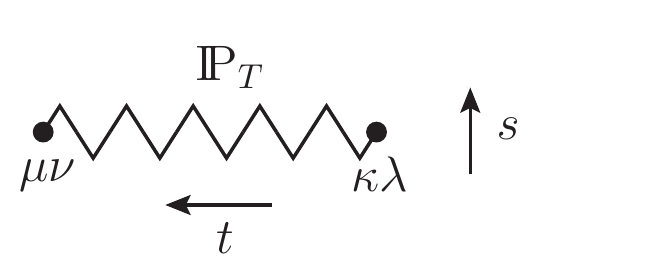} 
\begin{equation}\label{tensor_pomeron_propagator}
i\Delta^{(\Pom_{T})}_{\mu\nu,\kappa\lambda} (s,t) 
= \frac{1}{4s} \left(g_{\mu\kappa} g_{\nu\lambda} + g_{\mu\lambda} g_{\nu\kappa} 
- \frac{1}{2} g_{\mu\nu} g_{\kappa\lambda} \right) 
\, (-i s \alpha'_{\Pom})^{\alpha_{\Pom}(t)-1} \,.
\end{equation}
We assume a standard linear form for the pomeron trajectory (see \cite{Donnachie:2002en,Ewerz:2013kda})
\begin{equation}\label{pomeron_propagator_parameters}
\begin{split}
\alpha_{\Pom}(t) &= 1 + \epsilon_{\Pom}+ \alpha'_{\Pom} t \,,\\
\epsilon_{\Pom} &= 0.0808 \,,\\
\alpha'_{\Pom} &= 0.25 \,\mbox{GeV}^{-2} \,.
\end{split}
\end{equation}
%

\subsection{Vector pomeron $\Pom_{V}$}
\label{sec:Vectorpomeron}

Here the pomeron has a vector index and is coupled to a vector current
\begin{equation}\label{vector_pomeron_current}
\begin{split}
\mathcal{L}'_{V}(x) &= J_{V \mu}(x) \,\Pom_{V}^{\mu}(x) \,,\\
J_{V \mu}(x) &=
- 3 \beta_{\Pom NN} \, M_{0} \,
\bar{\psi}_{p}(x) \gamma_{\mu} \psi_{p}(x) \,,
\end{split}
\end{equation}
where $M_{0} \equiv 1$~GeV is introduced for dimensional reasons.
The corresponding $\Pom_{V}pp$ vertex and $\Pom_{V}$ propagator
are as follows (see (B.1) and (B.2) of \cite{Lebiedowicz:2013ika}): 
\newline
\hspace*{1.8cm}\includegraphics[width=145pt]{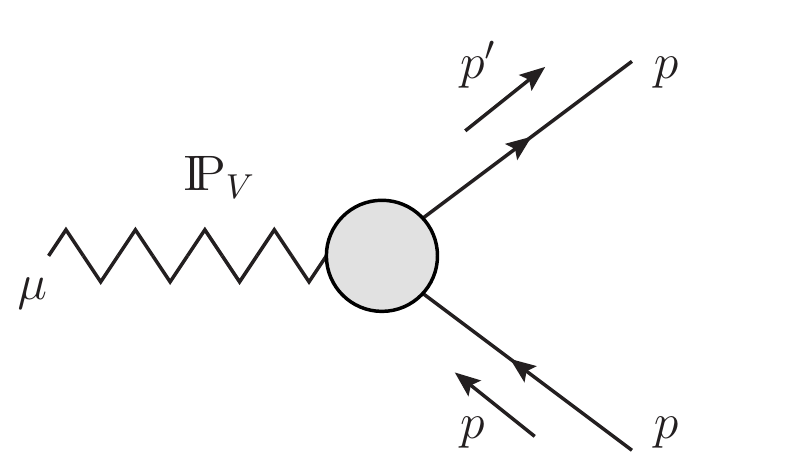}
\begin{equation}\label{vector_pomeron_vertex}
\begin{split}
i \Gamma_{\mu}^{(\Pom_{V} pp)}(p',p) =
-i \,3 \beta_{\Pom NN} M_0 F_{1}[(p'-p)^2] \, \gamma_{\mu} \,,
\end{split}
\end{equation}
\newline
\hspace*{1.8cm}\includegraphics[width=125pt]{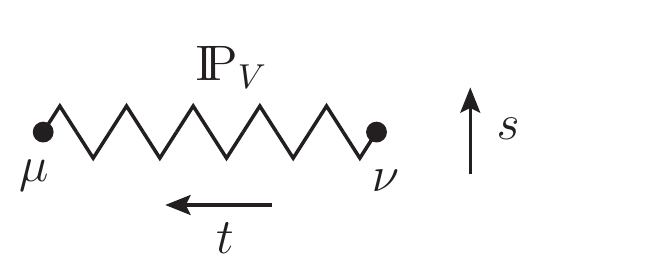} 
\begin{equation}\label{vector_pomeron_propagator}
\begin{split}
i\Delta^{(\Pom_{V})}_{\mu\nu} (s,t) 
= \frac{1}{M_{0}^{2}} g_{\mu\nu}
\, (-i s \alpha'_{\Pom})^{\alpha_{\Pom}(t)-1} \,.
\end{split}
\end{equation}
%

\subsection{Scalar pomeron $\Pom_{S}$}
\label{sec:Scalarpomeron}
Finally, for comparison, we discuss also a scalar pomeron $\Pom_{S}$ coupled 
to a scalar current
\begin{equation}\label{scalar_pomeron_current}
\begin{split}
\mathcal{L}'_{S}(x) &= J_{S}(x) \,\Pom_{S}(x) \,,\\
J_{S}(x) &=
- 3 \beta_{\Pom NN} \, M_{0} \,
\bar{\psi}_{p}(x) \psi_{p}(x) \,.
\end{split}
\end{equation}
Here we have as $\Pom_{S} pp$ coupling and as $\Pom_{S}$ propagator
the following expressions: 
\newline
\hspace*{1.8cm}\includegraphics[width=145pt]{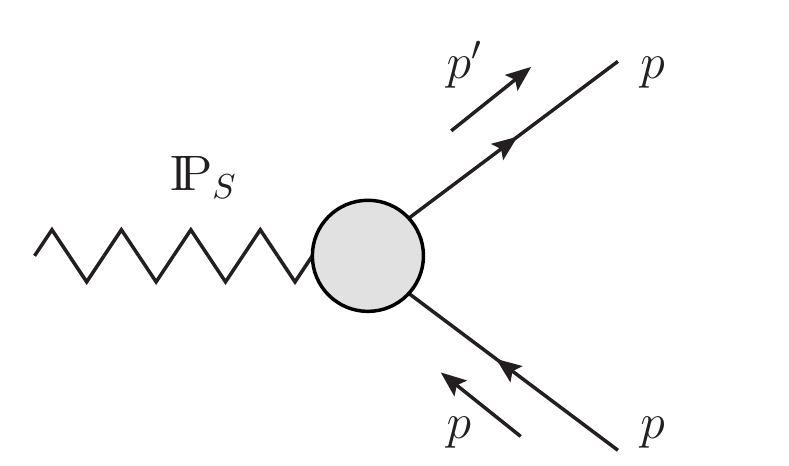}
\begin{equation}\label{scalar_pomeron_vertex}
\begin{split}
i \Gamma^{(\Pom_{S} pp)}(p',p) =
-i \,3 \beta_{\Pom NN} M_0 F_{1}[(p'-p)^2] \,, 
\end{split}
\end{equation}
\newline
\hspace*{1.8cm}\includegraphics[width=125pt]{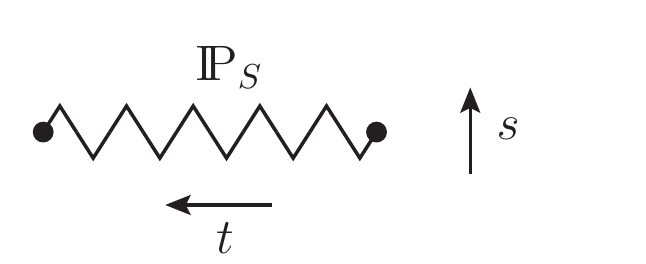} 
\begin{equation}\label{scalar_pomeron_propagator}
\begin{split}
i\Delta^{(\Pom_{S})}(s,t) 
= \frac{s}{2 m_{p}^{2} M_{0}^{2}}
\, (-i s \alpha'_{\Pom})^{\alpha_{\Pom}(t)-1} \,.
\end{split}
\end{equation}
%

\section{Helicity amplitudes}
\label{sec:Theamplitudes}
There are 16 helicity amplitudes for the reaction (\ref{pp_reaction}), 
defined as the $\mathcal{T}$-matrix elements 
\begin{equation}\label{general_amplitude}
\begin{split}
\Braket{p(p_{3},s_{3}),p(p_{4},s_{4})|{\cal T}|p(p_{1},s_{1}),p(p_{2},s_{2})}
&\equiv \Braket{2s_{3},2s_{4}|{\cal T}|2s_{1},2s_{2}} \,,\\
\;s_{j} \in \{1/2, -1/2\},& \quad j = 1, \dots, 4\,.
\end{split}
\end{equation}
The standard references for the general analysis of these amplitudes are 
\cite{Goldberger:1960md,Buttimore:1978ry,Buttimore:1998rj};
see also \cite{Leader:2001gr}.
Using rotational, parity ($P$), and time reversal ($T$) invariance,
and taking into account that protons are fermions one finds
that only five out of the 16 helicity amplitudes (\ref{general_amplitude})
are independent.
Conventionally these are taken as
\begin{equation}\label{five_amplitudes}
\begin{split}
&\phi_{1}(s,t) = \Braket{++|{\cal T}|++}\,,\\
&\phi_{2}(s,t) = \Braket{++|{\cal T}|--}\,,\\
&\phi_{3}(s,t) = \Braket{+-|{\cal T}|+-}\,,\\
&\phi_{4}(s,t) = \Braket{+-|{\cal T}|-+}\,,\\
&\phi_{5}(s,t) = \Braket{++|{\cal T}|+-}\,.
\end{split}
\end{equation}
The amplitudes with no helicity flip are $\phi_{1}$ and $\phi_{3}$,
with single flip $\phi_{5}$,
and with double flip $\phi_{2}$ and $\phi_{4}$.
Our normalisation is such that the differential
cross section for unpolarised protons is
\begin{equation}\label{dsig_dt}
\begin{split}
\frac{d\sigma(pp \to pp)}{dt} &= \frac{1}{16 \pi s (s-4m_{p}^{2})}
\frac{1}{4} \sum_{s_{1}, \dots, s_{4}} 
| \Braket{2s_{3},2s_{4}|{\cal T}|2s_{1},2s_{2}} |^{2}\\
& = \frac{1}{32 \pi} \frac{1}{s (s-4m_{p}^{2})}
\left\{
|\phi_{1}(s,t)|^{2}+|\phi_{2}(s,t)|^{2}+|\phi_{3}(s,t)|^{2}+|\phi_{4}(s,t)|^{2} \right.
\\
&\hspace*{3.4cm}
\left. 
+\, 4 \,|\phi_{5}(s,t)|^{2}
\right\}\,.
\end{split}
\end{equation}
The total cross section for unpolarised protons is\footnote{We note 
that the amplitudes $\phi_{j}^{\rm BGL}(s,t)$
defined in \cite{Buttimore:1978ry} are related to ours by 
$\phi_{j}^{\rm BGL}(s,t) = \phi_{j}(s,t)/(8 \pi)$. 
}
\begin{equation}\label{sigma_tot}
\begin{split}
\sigma_{\rm tot}(pp) 
& = \frac{1}{\sqrt{s (s-4m_{p}^{2})}}
    \frac{1}{4} \sum_{s_{1}, s_{2}} 
    {\rm Im} \Braket{2s_{1},2s_{2}|{\cal T}|2s_{1},2s_{2}} \mid_{t=0}\\
& = \frac{1}{2 \sqrt{s (s-4m_{p}^{2})}}
    \, {\rm Im} \left[ \phi_{1}(s,0)+\phi_{3}(s,0) \right]
\,.
\end{split}
\end{equation}

Now it is straightforward to calculate the amplitudes $\phi_{j}(s,t)$
in the tensor-, vector-, and scalar-pomeron models 
of section \ref{sec:Theoretical_framework}.
We use the phase conventions for the proton spinors of definite
helicity as given in (4.10) of \cite{Goldberger:1960md}
and find the following results. 

\paragraph{Tensor pomeron}
\begin{equation}
\begin{split}
\label{amp_tensor}
i\Braket{2s_{3},2s_{4}|{\cal T}|2s_{1},2s_{2}} =& \, 
i\Braket{p(p_{3},s_{3})|J_{T\mu \nu}(0)|p(p_{1},s_{1})} 
i\Delta^{(\Pom_{T})\mu\nu,\kappa\lambda}(s,t) 
\\
&{}
\times i\Braket{p(p_{4},s_{4})|J_{T\,\kappa\lambda}(0)|p(p_{2},s_{2})}\,,  \\
\Braket{p(p',s')|J_{T\mu \nu}(0)|p(p,s)} =& \, 
\bar{u}(p',s') \, \Gamma^{(\Pom_{T}pp)}_{\mu \nu}(p',p)\, u(p,s)\,.
\end{split}
\end{equation}
Inserting here the expressions from (\ref{tensor_pomeron_vertex})
and (\ref{tensor_pomeron_propagator})
we get the amplitudes $\phi_{j}(s,t)$.
It is convenient to pull out a common factor
\begin{equation}\label{F_factor}
\begin{split}
{\cal F}(s,t) 
&= i \left[ 3 \beta_{\Pom NN} F_{1}(t) \right]^{2}
   \frac{1}{4s} (-i s \alpha'_{\Pom})^{\alpha_{\Pom}(t)-1}\\
&= \left[ 3 \beta_{\Pom NN} F_{1}(t) \right]^{2}
   \frac{1}{4s} (s \alpha'_{\Pom})^{\alpha_{\Pom}(t)-1}
   \left[ \sin\left(\frac{\pi}{2}({\alpha_{\Pom}(t)-1}) \right)
     +i\cos\left(\frac{\pi}{2}({\alpha_{\Pom}(t)-1}) \right) \right]
\end{split}
\end{equation}
and to define reduced amplitudes by
\begin{equation}\label{reduced_amps}
\begin{split}
\hat{\phi}_{j}(s,t) = \phi_{j}(s,t)/{\cal F}(s,t)\,,
\qquad j = 1, \dots, 5.
\end{split}
\end{equation}
The results for $\hat{\phi}_{j}(s,t)$ in the tensor-pomeron model
are given in the column 'tensor' of Table~\ref{tab:table1}.
Terms of relative order $m_{p}^{2}/s$ and $|t|/s$ are neglected.
\begin{table}
\renewcommand{\baselinestretch}{1.2}\normalsize
\centering
\begin{tabular}{c|c|c|c|}
\cline{2-4}
\textbf{}                       
& \multicolumn{3}{c|}{pomeron ansatz}     \\ \cline{2-4} 
& tensor &   vector        &   scalar     \\ \hline
\multicolumn{1}{|c|}{$\hat{\phi}_{1}(s,t)$} 
& 8 $s^{2}$ & 8 $s^{2}$ & 8 $s^{2}$ \\ 
\multicolumn{1}{|c|}{$\hat{\phi}_{2}(s,t)$} 
& 10 $m_{p}^{2}t$ & 16 $m_{p}^{2}t$ & 2 $s^{2}t/m_{p}^{2}$          \\
\multicolumn{1}{|c|}{$\hat{\phi}_{3}(s,t)$} 
& 8 $s^{2}$ & 8 $s^{2}$ & 8 $s^{2}$ \\ 
\multicolumn{1}{|c|}{$\hat{\phi}_{4}(s,t)$} 
& -10 $m_{p}^{2}t$ & -16 $m_{p}^{2}t$ & -2 $s^{2}t/m_{p}^{2}$          \\ 
\multicolumn{1}{|c|}{$\hat{\phi}_{5}(s,t)$} 
& -8 $s m_{p} \sqrt{-t}$ & -8 $s m_{p} \sqrt{-t}$ & -4 $s^{2}\sqrt{-t}/m_{p}$ \\ \hline
\end{tabular}
\renewcommand{\baselinestretch}{1}\normalsize
\caption{Results for the reduced $pp$ scattering amplitudes $\hat{\phi}_{j}$
(\ref{reduced_amps}), $j = 1, \dots, 5$, for the tensor-, vector-, and
scalar-pomeron ans{\"a}tze.
Terms of relative order $m_{p}^{2}/s$ and $|t|/s$ are neglected. 
\label{tab:table1}
}
\end{table}

\paragraph{Vector pomeron}
\begin{equation}
\begin{split}
\label{amp_vector}
i\Braket{2s_{3},2s_{4}|{\cal T}|2s_{1},2s_{2}}=& \, 
i\Braket{p(p_{3},s_{3})|J_{V\mu}(0)|p(p_{1},s_{1})} \;
i\Delta^{(\Pom_{V})\mu\nu}(s,t)
\\
&{}
\times 
i\Braket{p(p_{4},s_{4})|J_{V\nu}(0)|p(p_{2},s_{2})}\,,\\
\Braket{p(p',s')|J_{V\mu}(0)|p(p,s)} =& \, 
\bar{u}(p',s') \, \Gamma^{(\Pom_{V}pp)}_{\mu}(p',p) \, u(p,s)\,.
\end{split}
\end{equation}
Inserting here the expressions from (\ref{vector_pomeron_vertex})
and (\ref{vector_pomeron_propagator})
we get the reduced amplitudes $\hat{\phi}_{j}(s,t)$ (\ref{reduced_amps})
in the column 'vector' of Table~\ref{tab:table1}. 

\paragraph{Scalar pomeron}
\begin{equation}
\label{amp_scalar}
\begin{split}
i\Braket{2s_{3},2s_{4}|{\cal T}|2s_{1},2s_{2}}=& \, 
i\Braket{p(p_{3},s_{3})|J_{S}(0)|p(p_{1},s_{1})}
i\Delta^{(\Pom_{S})}(s,t)
\\
&{}
\times i\Braket{p(p_{4},s_{4})|J_{S}(0)|p(p_{2},s_{2})}\,,\\
\Braket{p(p',s')|J_{S}(0)|p(p,s)} =& \, 
\bar{u}(p',s') \, \Gamma^{(\Pom_{S}pp)}(p',p) \, u(p,s)\,.
\end{split} 
\end{equation}
Inserting here the expressions from (\ref{scalar_pomeron_vertex})
and (\ref{scalar_pomeron_propagator})
we get the reduced amplitudes $\hat{\phi}_{j}(s,t)$ (\ref{reduced_amps})
in the column 'scalar' of Table~\ref{tab:table1}. 

\section{Discussion and comparison with experiment}
\label{sec:Discussion_and_comparison}

We note first that, by construction,
the non-flip amplitudes $\phi_{1}(s,t)$ and $\phi_{3}(s,t)$
are the same for all three pomeron hypotheses.
Thus, from (\ref{sigma_tot}), we get the same total cross section
\begin{equation}\label{sigma_tot_aux}
\begin{split}
\sigma_{\rm tot}(pp) = 2 \left( 3 \beta_{\Pom NN} \right)^{2}
\cos\left(\frac{\pi}{2} (\alpha_{\Pom}(0)-1) \right)
\left( s \alpha'_{\Pom} \right)^{\alpha_{\Pom}(0)-1}\,.
\end{split}
\end{equation}
This is the standard expression for the soft-pomeron contribution
to $\sigma_{\rm tot}(pp)$; see \cite{Donnachie:2002en} and (6.41) of \cite{Ewerz:2013kda}.

Now we consider the vector-pomeron case.
There, big problems arise if we consider $pp$ and $\bar{p}p$ scattering.
For $\bar{p}p$ elastic scattering 
we get an expression analogous to (\ref{amp_vector}), 
\begin{equation}\label{amp_vector_ppbar}
\begin{split}
&i\Braket{\bar{p}(p_{3},s_{3}), p(p_{4},s_{4})|{\cal T}|
\bar{p}(p_{1},s_{1}), p(p_{2},s_{2})}\\
&=
i\Braket{\bar{p}(p_{3},s_{3})|J_{V\mu}(0)|\bar{p}(p_{1},s_{1})} \;
i\Delta^{(\Pom_{V})\mu\nu}(s,t)\;
i\Braket{p(p_{4},s_{4})|J_{V\nu}(0)|p(p_{2},s_{2})}\,.
\end{split}
\end{equation}
A charge-conjugation transformation $C$ gives
\begin{equation}\label{amp_vector_ppbar_aux1}
\begin{split}
 \Braket{\bar{p}(p_{3},s_{3})|J_{V\mu}(0)|\bar{p}(p_{1},s_{1})} =
-\Braket{p (p_{3},s_{3})|J_{V\mu}(0)|p (p_{1},s_{1})}\,,
\end{split}
\end{equation}
as follows from the standard $C$-transformation rules of 
the (anti)proton states and of the bilinear expression for the vector 
current in terms of the proton fields \eqref{vector_pomeron_current}. 
Thus, we get for a vector pomeron
\begin{equation}\label{amp_vector_ppbar_aux2}
\begin{split}
&\Braket{\bar{p} (p_{3},s_{3}), p(p_{4},s_{4})|{\cal T}|
\bar{p}(p_{1},s_{1}), p(p_{2},s_{2})}
\\
&\hspace*{.5cm}
= - \Braket{p(p_{3},s_{3}), p(p_{4},s_{4})|{\cal T}|
p(p_{1},s_{1}), p(p_{2},s_{2})}
\end{split}
\end{equation}
and hence from (\ref{sigma_tot})
\begin{equation}\label{sig_tot_ppbar}
\begin{split}
\sigma_{\rm tot}(\bar{p}p)= - \sigma_{\rm tot}(pp)
\,.
\end{split}
\end{equation}
Clearly, this result does not make sense for a non-vanishing cross section 
and would contradict the rules of quantum field theory. 
Thus, we shall not consider a vector pomeron any further.\footnote{Let us 
remark here, however, that for proton-proton scattering the parameter 
$r_5$ discussed below would be the same in the vector-pomeron 
case as in the tensor-pomeron case as can be inferred from the amplitudes 
in Table~\ref{tab:table1}.} 

We are left with the tensor- and scalar-pomeron hypotheses.
We note that the $C$ transformation here gives 
(see (\ref{tensor_pomeron_current}) and (\ref{scalar_pomeron_current}))
\begin{equation}\label{amp_tensor_aux}
\begin{split}
\Braket{\bar{p}(p_3,s_3)|J_{T\mu \nu}(0)|\bar{p}(p_1,s_1)} =
\Braket{p (p_3,s_3)|J_{T\mu \nu}(0)| p (p_1,s_1)}\,,
\end{split}
\end{equation}
and
\begin{equation}\label{amp_scalar_aux}
\begin{split}
\Braket{\bar{p}(p_3,s_3)|J_{S}(0)|\bar{p}(p_1,s_1)} =
\Braket{p (p_3,s_3)|J_{S}(0)| p (p_1,s_1)}\,.
\end{split}
\end{equation}
This implies in both cases the equality of the pomeron contributions
to the $pp$ and $\bar{p}p$ scattering amplitudes,
as should be the case. 

In order to discriminate between the tensor and scalar pomeron cases
we turn to data from the STAR experiment at RHIC \cite{Adamczyk:2012kn}.
There, a measurement of the ratio
of single-flip to non-flip amplitudes at $\sqrt{s} = 200 \,\mbox{GeV}$
was performed.
The relevant quantity is
\begin{equation}\label{r5}
\begin{split}
r_{5}(s,t) = 
\frac{2m_{p} \,\phi_{5}(s,t)}{\sqrt{-t} \,
{\rm Im}\left[ \phi_{1}(s,t) + \phi_{3}(s,t)\right]}\,.
\end{split}
\end{equation}
From \eqref{F_factor}, \eqref{reduced_amps} and Table~\ref{tab:table1}
we find for the tensor pomeron
\begin{equation}\label{r5_tensor}
\begin{split}
r_{5}^{\Pom_{T}}(s,t) = -\frac{m_{p}^{2}}{s}
\left[ i + \tan\left(\frac{\pi}{2} (\alpha_{\Pom}(t)-1)\right) \right]\,.
\end{split}
\end{equation}
For the scalar pomeron, on the other hand, we get 
\begin{equation}\label{r5_scalar}
\begin{split}
r_{5}^{\Pom_{S}}(s,t) = -\frac{1}{2}
\left[ i + \tan\left(\frac{\pi}{2} (\alpha_{\Pom}(t)-1)\right) \right]\,.
\end{split}
\end{equation}
The measurement of $r_{5}$ in \cite{Adamczyk:2012kn} 
is done for $0.003 \leqslant |t| \leqslant 0.035$~GeV$^{2}$ 
and no $t$-dependence of $r_5$ is observed in this range. 
The latter observation is in agreement with our results 
\eqref{r5_tensor} and \eqref{r5_scalar} which also imply only 
a weak $t$-dependence of $r_5(s,t)$. 
Therefore, we can approximately set $t=0$
in (\ref{r5_tensor}) and (\ref{r5_scalar})
and obtain with $\sqrt{s} = 200 \,\mbox{GeV}$ 
\begin{eqnarray}
&&r_{5}^{\Pom_{T}}(s,0) =
\left( -0.28 - i\, 2.20 \right) \times 10^{-5}\,,
\label{r5_tensor_result}\\
&&r_{5}^{\Pom_{S}}(s,0) =
-0.064 - i\, 0.500\,.
\label{r5_scalar_result}
\end{eqnarray}
It is worth pointing out that in the high-energy limit ($s \gg m_p^2$) 
these results have very small remaining uncertainties since only the 
pomeron intercept $\alpha_{\Pom}(0)$ enters which is rather well 
determined experimentally. However, as the result \eqref{r5_tensor_result} 
for the tensor pomeron is very small in absolute terms, it is conceivable 
that in this case subleading terms might 
be relevant for a calculation of $r_5$ with very high precision. But as we 
will see momentarily, the current experimental uncertainty does not 
allow a determination of $r_5$ to that precision anyway. 

\begin{figure}
\begin{center}
\includegraphics[width=0.6 \textwidth]{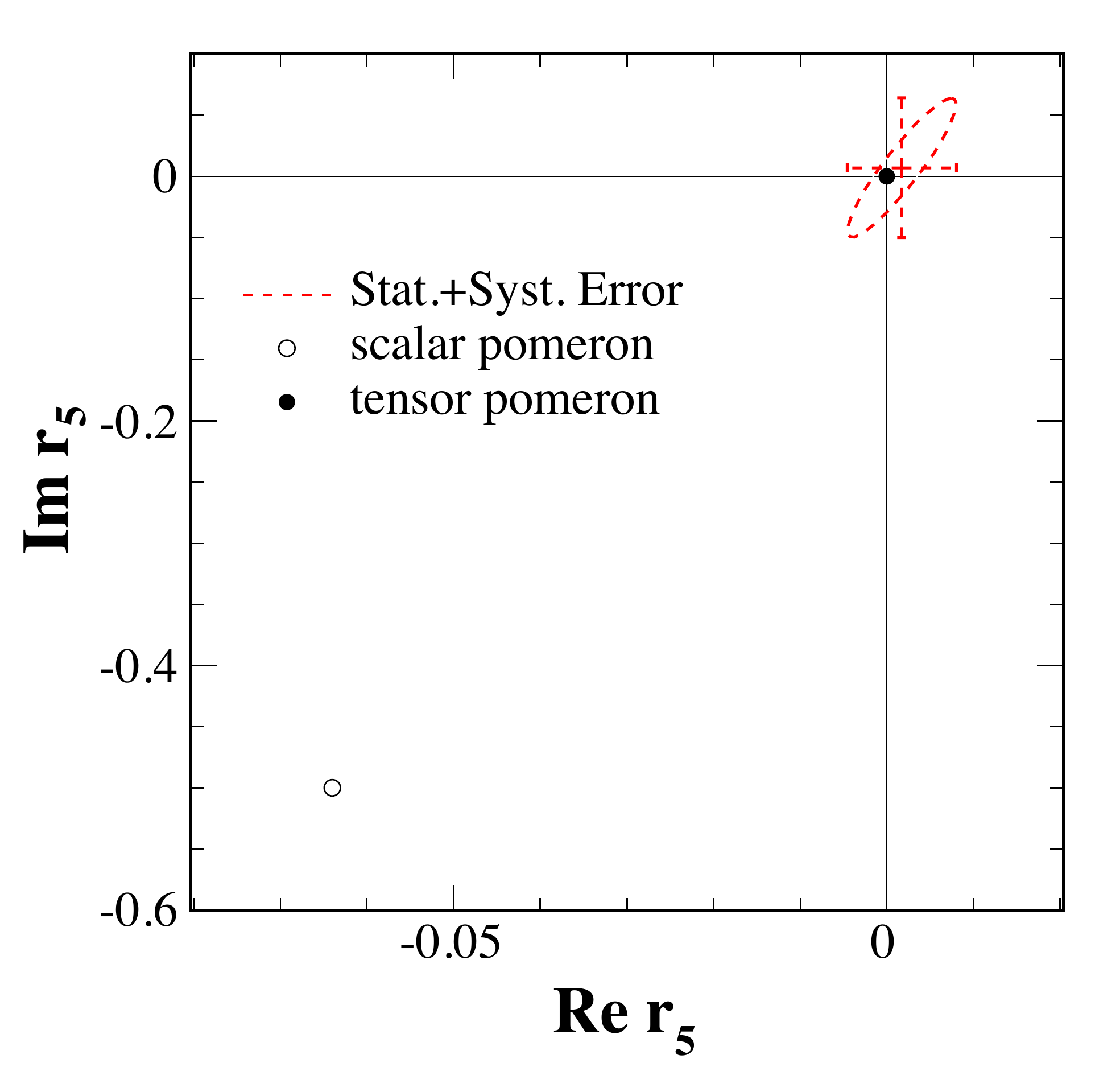}\\ \vspace*{.4cm}
\includegraphics[width=0.6 \textwidth]{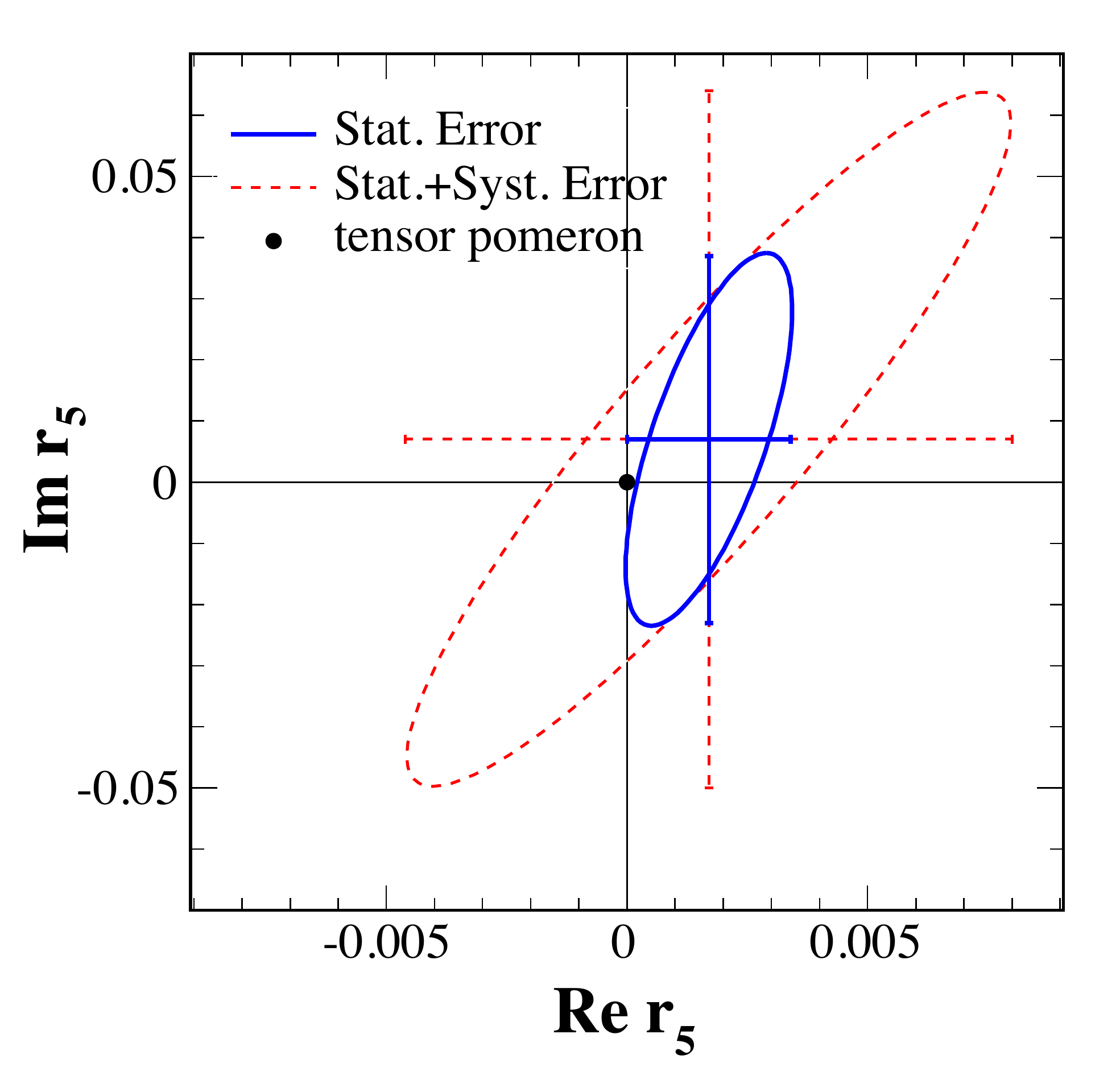}       
\caption{
The experimental results for $r_{5}$ at $\sqrt{s} = 200\, \mbox{GeV}$ 
from Fig.~5 of \cite{Adamczyk:2012kn} together with our results for
the tensor and the scalar pomeron; 
see (\ref{r5_tensor_result}) and (\ref{r5_scalar_result}). 
The first figure shows the experimental data together with the results for 
both pomeron models. The second figure shows a magnified view 
of the relevant region around zero and contains only the clearly favoured 
tensor-pomeron point. 
\label{fig:figure2}
}
\end{center}
\end{figure}

In Fig.~\ref{fig:figure2} we show the experimental result for $r_5$ 
of \cite{Adamczyk:2012kn} (as given in Fig.~5 there) together with our results 
(\ref{r5_tensor_result}) and (\ref{r5_scalar_result}).
Clearly, the tensor-pomeron result is perfectly
compatible with the experiment. The scalar-pomeron result, on the other 
hand, is far outside the experimental error ellipse. The tensor pomeron 
is hence strongly favoured by the data, while the scalar pomeron is ruled out. 

\section{Conclusions}
\label{sec:Conclusions}

In this article we have confronted three hypotheses for the soft pomeron -- 
tensor, vector, and scalar -- with experimental data on polarised high-energy $pp$ 
elastic scattering from the STAR collaboration \cite{Adamczyk:2012kn}. 
Studying the ratio $r_5$ of single-helicity-flip to non-flip amplitudes 
we found that the STAR data are consistent with a tensor pomeron while 
they clearly exclude a scalar pomeron. We have further argued that a vector pomeron 
assumption is in contradiction to the rules of quantum field theory. 
We therefore conclude that the tensor pomeron is the only viable option.

Attempts to relate the pomeron to tensors were, in fact, already discussed 
in the 1960's \cite{Freund:1962,Freund:1971sh,Carlitz:1971ee}. In \cite{Freund:1962} 
the energy-momentum tensor was considered and from that and some further 
assumptions the ratio of meson-baryon to baryon-baryon cross sections 
was obtained giving, for instance, 
\begin{equation}
\label{ratiofromfreund}
\frac{\sigma_{\rm tot}(\pi p)}{\sigma_{\rm tot}(pp)} \approx \frac{m_K}{m_B} \approx \frac{1}{2} \,.
\end{equation}
Here $m_K$ and $m_B$ are a mean meson mass, taken as the one of the $K$ meson, 
and a mean baryon mass, respectively. 
This does not work phenomenologically, as in fact 
$\sigma_{\rm tot}(\pi p)/\sigma_{\rm tot}(pp) \approx 2/3$. 
Also, we cannot see the physics which would make the $\pi p$ cross section 
proportional to the $K$ mass. 

In this connection we may discuss the consequences of the 
hypothesis that the tensor current $J_{T\mu\nu}(x)$ 
in \eqref{tensor_pomeron_current} is proportional to the energy-momentum 
tensor with a {\em universal} constant of proportionality, 
independent of the hadron considered. It is easy to see that this leads 
to the same pomeron part of the total cross section for all hadron-hadron 
scatterings, for instance $\sigma(pp) = \sigma(\pi p) = \sigma(J/\psi \, p)$. 
Clearly, also this does not work phenomenologically. Our $J_{T\mu\nu}(x)$ 
in \eqref{tensor_pomeron_current} {\em cannot} be universally proportional 
to the energy-momentum tensor. 

In \cite{Freund:1971sh,Carlitz:1971ee} attempts were made to relate the 
pomeron properties, in particular its couplings to hadrons, to those of 
tensor mesons of the $q\bar{q}$ type. However, with the advent of QCD 
and gluons it has become clear that the pomeron is a predominantly 
gluonic object; see the pioneering papers \cite{Low:1975sv,Nussinov:1975mw}. 
Thus, if one wants to relate the pomeron properties to mesonic ones 
it is natural to look for glueballs, and this, indeed, has been and is being 
done frequently. Here, one problem is that even today the status 
of glueballs is not particularly clear; for a review see \cite{Ochs:2013gi}. 
A vast literature exists dealing with the pomeron in perturbative QCD, 
starting from the celebrated work \cite{Kuraev:1977fs,Balitsky:1978ic}. 
Questions similar to those addressed in the present paper for the soft pomeron 
could be interesting also in the context of the perturbative pomeron, 
but this would be beyond the scope of the present paper. 

For the soft pomeron phenomenology, for a long time then, 
a sort of vector pomeron was commonly used,
following \cite{Donnachie:1983hf,Donnachie:1985iz,Donnachie:1987gu}, 
although it was clear that this could not be
completely correct due to the problems with charge conjugation
explained in section~\ref{sec:Discussion_and_comparison}. 
A first attempt to understand the soft pomeron in the framework 
of a toy model of nonperturbative QCD was made in \cite{Landshoff:1986yj}. 
In \cite{Nachtmann:1991ua} functional integral techniques were used to analyse
high-energy soft hadron-hadron scattering in QCD in a nonperturbative framework. 
It was shown in chapter~6 of \cite{Nachtmann:1991ua}
that the pomeron exchange can be understood as a coherent sum
of exchanges of spin $2+4+6+\dots$. 
Going through the arguments there one can see that
basically this structure is due to the helicity
conserving fundamental quark-gluon coupling in QCD. 
In \cite{Ewerz:2013kda} a tensor pomeron was introduced which again can be
viewed as a coherent sum of exchanges of spin $2+4+6+\dots$ 
(see appendix~B of \cite{Ewerz:2013kda})
thus making contact with the considerations in QCD of \cite{Nachtmann:1991ua}.
We note that writing a regge exchange as a coherent sum 
of elementary spin exchanges goes back to \cite{vanHove:1967zz}.
Concerning specific experimental tests
for the spin structure of the pomeron
we should mention \cite{Arens:1996xw}
where such tests were proposed for diffractive
deep inelastic electron-proton scattering.
Similar techniques were proposed in \cite{Close:1999is,Close:1999bi}
for central production of meson resonances in $pp$ collisions
\begin{eqnarray}\label{pp_pMp}
p + p \longrightarrow p + {\rm meson} + p \,.
\end{eqnarray}
In the light of our discussion here 
we cannot support the conclusions of \cite{Close:1999is,Close:1999bi}
that the pomeron couples like a non-conserved vector current.
In \cite{Lebiedowicz:2013ika,Lebiedowicz:2014bea,Lebiedowicz:2016ioh} 
the question of central production
(\ref{pp_pMp}) was taken up again from the point of view
of the tensor pomeron and it was shown that this does quite well
in reproducing the data where available.
It turned out, however, that central production with
pomeron-pomeron fusion,
\begin{eqnarray}\label{pompom_M}
\Pom + \Pom \longrightarrow {\rm meson}  \,,
\end{eqnarray}
was not too sensitive to the nature of the pomeron,
tensor or vector. But central production
with fusion of a $C = -1$ object with the pomeron, e.\,g.
\begin{eqnarray}\label{gampom_pipi}
\gamma + \Pom \longrightarrow \pi^{+} + \pi^{-}
\end{eqnarray}
is extremely sensitive to the nature of pomeron.
For a tensor (vector) pomeron the $\pi^{+}\pi^{-}$ pair
in (\ref{gampom_pipi}) is in an antisymmetric (symmetric)
state under the exchange $\pi^{+} \leftrightarrow \pi^{-}$.
Needless to say that since the pomeron has
$C = +1$ the $\pi^{+}\pi^{-}$ pair in (\ref{gampom_pipi})
must be in an antisymmetric state. Thus also from this 
point, a vector pomeron is excluded. 
Finally, also investigations of the pomeron using the 
AdS/CFT correspondence prefer a tensor nature for pomeron 
exchange \cite{Domokos:2009hm,Iatrakis:2016rvj}.

According to the results for the helicity-amplitudes in 
polarised high-energy $pp$ elastic scattering presented in this work 
the soft pomeron should be described as a rank-2 symmetric tensor exchange, 
as for example in the model of \cite{Ewerz:2013kda}. 
It is not a priori clear that the pomeron exhibits the same spin 
structure also in reactions involving high momentum transfers, i.\,e.\ 
reactions with the exchange of a hard (perturbative) pomeron. 
We would find it particularly desirable  
to study the spin structure of the pomeron in the interesting transition 
region between soft and hard reactions. It would therefore 
be useful to discuss further observables that are sensitive to the 
spin structure of the pomeron and that can be experimentally studied 
in a wide range of kinematic regimes. 

\section*{Acknowledgments}
The authors would like to thank W.~Guryn for giving 
an inspiring talk on the STAR experiment
at the meeting ''Diffractive and Electromagnetic
Processes at High Energies`` at Bad Honnef in 2015 and for 
providing the data from \cite{Adamczyk:2012kn}. 
This research was partially supported by 
the MNiSW Grant No.~IP2014~025173,
the Polish National Science Centre Grant No.~DEC-2014/15/B/ST2/02528, 
and by the Centre for Innovation and Transfer of Natural Sciences 
and Engineering Knowledge in Rzesz\'ow.

\end{document}